\documentclass[structabstract]{aa}  
\usepackage{graphicx}
\usepackage{lscape}
\usepackage{txfonts}
\usepackage{natbib}
\usepackage{sidecap}
\usepackage{comment}
\usepackage{color}
\usepackage{booktabs}

\begin{document}
\title{Discovery of a young and massive stellar cluster:}
\subtitle{Spectrophotometric near-infrared study of Masgomas-1}
\author{S. Ram\'irez Alegr\'ia\inst{1,2} \and A. Mar\'in-Franch\inst{3,4} \and A. Herrero\inst{1,2}}

\institute{Instituto de Astrof\'isica de Canarias, 38205 La Laguna, Tenerife, Spain. \email{sramirez@iac.es, ahd} \and Departamento de Astrof\'isica, Universidad de La Laguna, E-38205 La Laguna, Tenerife, Spain. \and Centro de Estudios de F\'isica del Cosmos de Arag\'on (CEFCA), E-44001, Teruel, Spain. \email{amarin@cefca.es} \and Departamento de Astrof\'isica, Universidad Complutense de Madrid, E-38040, Madrid, Spain. \\}
\date{Received 2011 / Accepted 2011}

\abstract
{Recent near-infrared data have contributed to the discovery of new (obscured) 
massive stellar clusters and massive stellar populations in previously
 known clusters in our Galaxy. These discoveries lead us to view the 
Milky Way as an active star-forming machine.}
{The main purpose of this work is to  determine physically the main parameters
(distance, size, total mass and age) of Masgomas-1, the first massive cluster
discovered by our systematic search programme.}
{Using near-infrared ($J$, $H$, and $K_S$) photometry we selected 23 OB-type and
five red supergiant candidates for multi-object $H$- and $K$-spectroscopy and spectral
classification.}
{Of the 28 spectroscopically observed stars, 17 were classified as OB-type, four as 
supergiants, one as an A-type dwarf star, and six as late-type giant stars. The presence
of a supergiant population implies a massive nature of Masgomas-1, supported
by our estimate of the cluster initial total mass of $(1.94\pm0.28)\cdot10^4 {M}_{\odot}$, obtained after integrating
the cluster mass function. The distance estimate of $3.53^{+1.55}_{-1.40}$ kpc locates the cluster closer than the 
Scutum--Centaurus base but still within that Galactic arm.  The presence of an O9\,V star and red supergiants in the
same population indicates that the cluster age is in the range of 8 to 10 Myr.}
{}
\keywords{Infrared: stars - Stars: early-types, supergiants, massive - Techniques: photometric, spectroscopic.}
 \titlerunning{Near-infrared study of Masgomas-1}
\maketitle

\section{Introduction}

 Massive stars are key components in the Galactic evolution. They change the star formation rate,
ionize their surrounding media with their winds, strengthen the formation of other stars, or 
deplete their native clouds through their own birth. They are short-lived objects, and their impact
on the interstellar medium also occurs on short times scales, playing a crucial role in the 
energy balance, dynamics, and chemical evolution of their host galaxy. We often find them embedded 
in obscured massive clusters characterized by high-extinction regions. Unfortunately, owing to this strong 
extinction, the optical detection of these objects is difficult, but they can be detected and observed 
using near-infrared instrumentation. All-sky surveys such as 2MASS \citep{skrutskie06}, GLIMPSE 
\citep{benjamin03}, and UKIDSS \citep{lawrence07} have allowed the discovery of the most massive 
stellar clusters in the Galaxy. Nevertheless, the census of massive clusters is far from complete; up to 
100 clusters with a mass greater than $10^4$ $M_{\odot}$ may remain hidden \citep{hansonpopescu08}. 
Systematic search programmes for these objects are necessary for a full understanding of our Galaxy.

 The MASGOMAS project \citep{marin09} has recently started a systematic search 
for massive cluster candidates in the Galactic disc. In a quick exploration using our search 
algorithm we have found a new massive cluster candidate, located near the direction of the 
Scutum--Centaurus arm base, and previously unidentified in the literature.
 
  In this exciting region, where the Scutum--Centaurus arm meets the Galactic central bar, several red supergiant
clusters have been found: \object{RSGC 1} \citep{figer06,davies08}, \object{RSGC 2} \citep{davies07},
\object{RSGC 3} \citep{clark09a}, and \object{Cl Alicante 8} \citep{negueruela10}. These massive 
clusters (each of them having a total mass estimate greater than $10^4$ $M_{\odot}$) are remarkable because of 
their population of red supergiant (RSG) stars, ranging between 8 and 26 RSGs. Their presence within 
a concentrated area in the Galactic plane ($l = 24\degr-29\degr$) and a similar distance ($d\sim 6$ kpc), makes
this region extremely interesting for searching massive cluster candidates.
 
\begin{figure*}
\centering
\includegraphics[width=2.3in,angle=90]{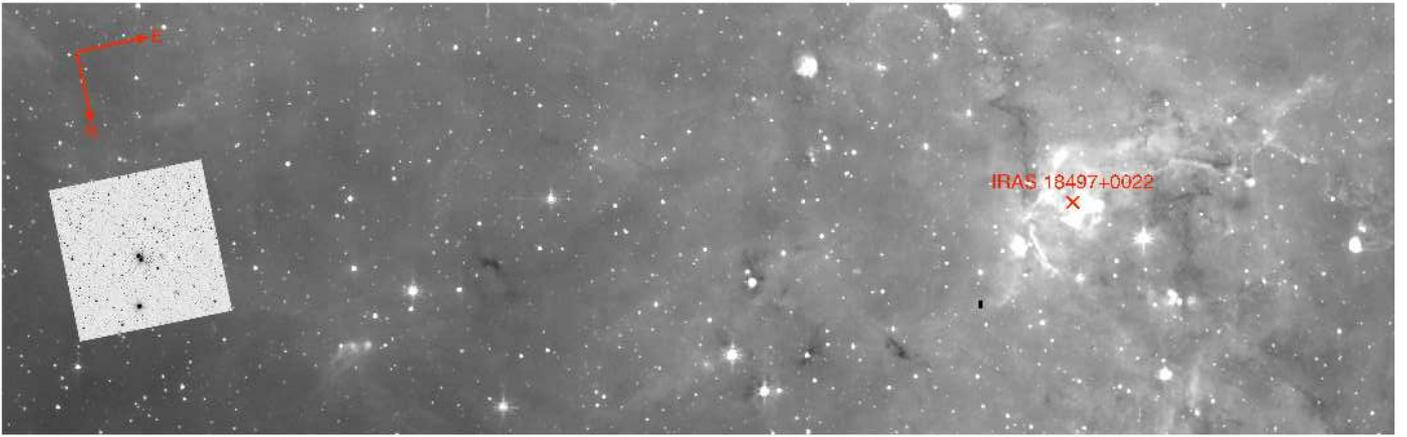} 
            \caption{Spitzer 8.0 $\mu m$ image section for the region around IRAS 18497+0022. The Spitzer section
            covers an area of 15$\times$50 square arc minutes and includes Masgomas-1, shown as observed with LIRIS
            on the left side of the image. The length of the orientation arrows is 2.5 arcmin.}
       \label{msgms01_large}
\end{figure*}

 The first new massive cluster candidate discovered by our group, and labelled as Masgomas-1, 
 is located in the Galactic plane near the Scutum--Centaurus base ($l = 33\,\fdg112$, $b = +0\,\fdg42$, and 
${\alpha}_{2000}=18^{\mathrm {h}}50^{\mathrm {m}}15^{\mathrm {s}}$, ${\delta}_{2000}=+00\degr21\arcmin04\arcsec$).
Masgomas-1 lies near the IRAS source \object{IRAS 18497+0022}, a very bright object in the Spitzer 8.0 $\mu m$ image, 
which extends over $\sim0.35\degr$ (see Fig. \ref{msgms01_large}). 
In the surroundings of Masgomas-1 we find two other IRAS sources (\object{IRAS 18476+0017}, \object{IRAS 18476+0019}), and
 the radio source \object{GPSR 033.086+0.434}. An inspection of 2MASS photometry of the smaller 
area within the square in Fig. \ref{msgms01_large} reveals the presence of three bright and red stars ($K_S <  6$ and $J-K_S \sim 4$), 
separated by $\sim$ 2 mag in $K_S$ from the other field stars. This characteristic has been observed previously in 
the red supergiant population of RSGC 1, 2, 3, and Alicante 8, making these three Masgomas-1 stars red supergiant candidates.

 In this article we describe our systematic search for massive stellar candidates based on the detection of 
OB-type star candidate over-densities (Section \ref{candidate_selection}) 
and spectrophotometric observations of Masgomas-1 (Section \ref{observations}). In Section \ref{res} we describe the 
near-infrared photometry of the cluster, the spectral classifications of the 28 spectroscopically 
observed stars, and the proper motion measures for 20 of them. An analysis of the cluster 
physical parameters (distance, extension, mass, and age) is given in Section \ref{discussion}. Our main 
results are summarized in Section \ref{conclusions}.

\section{Systematic search of obscured clusters and candidate selection \label{candidate_selection}}

For our preliminary cluster search we have used 2MASS photometry \citep{skrutskie06}, 
focused on Galactic coordinates in the ranges $l = 20^{\circ}-45^{\circ}$ and $b = 0^{\circ}-5^{\circ}$. 
This area was divided into smaller regions of 5$\times$5 square degrees. Only stars with 
$J$-, $H$-, and $K_S$-bands and good data quality (defined by photometric errors less than 0.1 and
quality flags equal to ``AAA'') were used in the candidate search.

 To select OB-type star candidates, we have filtered the photometry using the following criteria:
 \begin{enumerate}
 \item $K_S$ magnitude less than 12: By defining this limiting magnitude we aimed to prevent Poisson 
 noise, derived from the field stellar distribution, from prevailing over the candidate over-densities.
 \item Red $(J-K_S)$ colours: Unreddened foreground disc stars exhibit bluer colours in the CMD. To eliminate most of these stars
 from our systematic search, we applied a cut in $(J-K_S)$, discarding objects with $(J-K_S)< 0.5$.
 \item Reddening-free parameter between $-0.3$ and 0.3: The reddening-free parameter $Q_{IR}$, described by \citet{comeron05}
and \citet{negueruela07}, has a value close to zero for OB stars. Considering photometry errors we define a range for 
$Q_{IR}$ where OB-star candidates would be expected.  However, because A- and early F-type stars can mimic the $Q_{IR}$
 of an OB star, contamination by these stellar types is plausible.
 \end{enumerate}
 
 Using the sources that fulfil the previous criteria, which assures a selection of OB-star candidates,
we looked for over-densities of these stars to identify them as massive cluster candidates. One of these cluster 
candidates, without previous identification in the literature, is Masgomas-1.

\begin{figure} 
\centering
\includegraphics[width=3.9in]{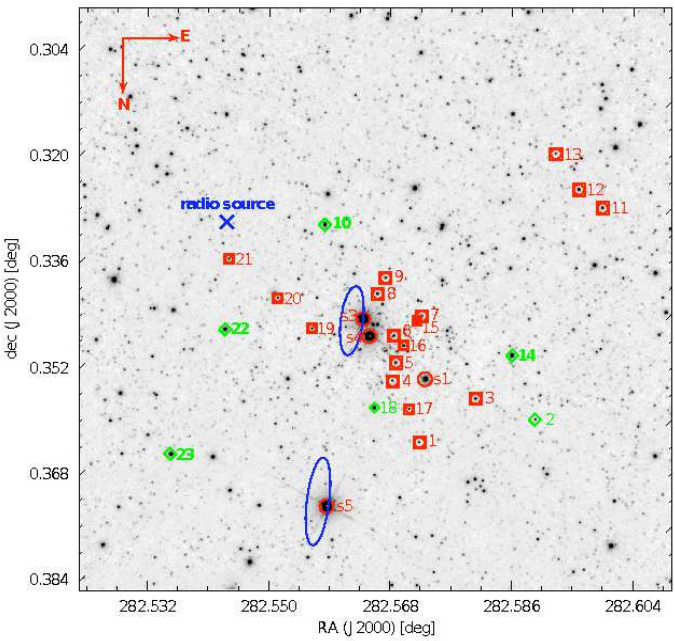}
            \caption{LIRIS $K_S$ image of Masgomas-1; the stars observed
            spectroscopically with the multi-object masks are shown with red squares (early-type stars), 
            and green diamonds (A- and late-type disc stars). Bright stars observed with long-slit spectroscopy
            are marked with red circles. The position of star s02, observed with long-slit, is not shown in the 
            figure because it lies 30 arcsec to the right of the field. Blue ellipses correspond to the 1$\sigma$ 
            positional uncertainties for the \object{IRAS 18476+0017} (central) and \object{IRAS 18476+0019} 
            (lower) sources. The (blue) cross shows the position of the radio source \object{GPSR 033.086+0.434}.}
       \label{mos_stars}
\end{figure}

 For Masgomas-1 most of the OB-type star candidates detected as an overdensity using the near-infrared 
data does not appear in optical images of the region. Even the central bright stars marked with red circles in Fig.
\ref{mos_stars} remains undetected in optical bands. The overdensity is concentrated within a $\sim1\,\farcm5$
radius and triplicates the number of OB-type candidates found in a control ring, which is concentric and covers same
area as the central search area.

\section{Observations of Masgomas-1 \label{observations}}

We based our study of Masgomas-1 on broad-band near-infrared imaging ($J$, $H$, $K_S$), medium 
resolution multi-object spectroscopy ($H$ and $K$) and long-slit spectroscopy ($H$ and $K$). All 
data were acquired with LIRIS, a near-infrared imager/spectrograph mounted at the Cassegrain focus 
of the 4.2 m William Herschel Telescope (Roque de Los Muchachos Observatory, La Palma). A summary
of the observations is given in Table \ref{tabla_obs}.

\begin{table*}
\caption{Summary of imaging and spectroscopic observations}

\begin{center}
\begin{tabular}{ccccc}
\toprule
Observation mode & Date & Filter & Exp. Time & Seeing \\
                                  &           &           &        [s]        &   [$''$]    \\
\midrule
Masgomas-1 field imaging   	&  2010-06-23 	&  $J$   	& 108.0	&  0.80  \\
                                                     	&  2010-06-23 	&  $H$  	&  36.0     	&  0.76  \\
   						&  2010-06-23	& $K_S$	&  36.0    	&  0.69  \\
\midrule
Multi-object spectroscopy (R$\sim$2500), mask A 	&  2010-06-23	& $H$	&  4800.0	&  0.76  \\
 										&  2010-06-24 	& $K$	&  4800.0	&  0.80    \\
\midrule
Multi-object spectroscopy (R$\sim$2500), mask B 	&  2010-06-25	& $H$	&  1920.0	&  0.84   \\
\midrule
Long-slit spectroscopy (R$\sim$2500) 			&  2010-06-23 	& $H$	&    720.0	&  0.64   \\
 										&  2010-06-24 	& $H$  	&    480.0 	&  0.56   \\
 										&  2010-06-24	& $K$  	&    320.0  &  0.60   \\
 										&  2011-09-15	& $H$  	&    420.0  &  0.55   \\
 										&  2011-09-15	& $K$  	&    260.0  &  0.72   \\
\bottomrule
\end{tabular}
\end{center}
\label{tabla_obs}
\end{table*}

\subsection{Imaging}
 
 The cluster candidate was observed on 2010 June 23 using LIRIS, an infrared camera equipped with a
 Hawaii 1024$\times$1024 HgCdTe array detector, with a field of view of $4.2'\times4.2'$ and a spatial scale of  
  $0\,\farcs25\mathrm{~pixel}^{-1}$. Images were obtained for filters $J$ (${\lambda}_C=1{.}250~\mu m$, $\Delta
 \lambda=0{.}160~\mu m$), $H$ (${\lambda}_C=1{.}635~\mu m$, $\Delta\lambda=0{.}290~\mu m$) and ${K}_{S}$ 
 (${\lambda}_C=2{.}150~\mu m$, $\Delta\lambda=0{.}320~\mu m$), with seeing values between $0\,\farcs69$ and 
 $0\,\farcs80$.
 
  To improve cosmic-ray and bad-pixel rejection, and to construct the sky image for subtraction, we observed on 
 a nine-point dithering mode. Data reduction (bad-pixel mask, flat correction, sky subtraction and alignment) was 
 made with FATBOY \citep{eikenberry06} and geometrical distortions were corrected with the LIRIS 
 reduction package LIRISDR\footnote{http://www.iac.es/project/LIRIS}. The final $K_S$ image for 
 Masgomas-1 is shown in Figure \ref{mos_stars}. In the $K_S$ image, as well as for the $J$ and $H$ 
images, three bright stars stand out, two of them in the centre of the field (s03, s04 and s05). These 
two central bright stars, observed with long-slit spectroscopy as described in next section, set the centre 
of our cluster candidate. We also marked in the image the position of the spectroscopically observed 
OB-type candidates. Most of the candidates are concentrated around the aforementioned 
bright central stars, and the central coordinates of the OB-type candidates overdensity coincide with the 
position of the bright central stars.

 Instrumental photometry was made with DAOPHOT II, ALLSTAR and ALLFRAME \citep{stetson94}. The photometry was
 cleaned of non-stellar and poorly measured objects using ALLFRAME sharp index ($-0.25 < sharp < 0.25$) and 
 PSF fitting $\sigma$ (less than 0.1). The photometric calibration for the three filters was performed using sources from 
 the 2MASS catalogue. We selected 362 isolated and non-saturated stars with photometry in 2MASS, well distributed 
 over the images and the $(J-K_S)-K_S$ colour-magnitude diagram. For stars with $K_S$ less than 9 mag, we adopted 
 the 2MASS photometry, due to saturation.

 The astrometrical calibration was performed with SKYCAT, matching physical and equatorial coordinates for 14 
 sources that are well-distributed in our LIRIS image. Typical values for the fitting RMS were less than 0.15 arcsec 
 for the three bands, which was adequate for the mask design requirement.

\subsection{Infrared spectroscopy}
 The near-infrared spectra were obtained on 2010 June 23, 24, 25, and 2011 September 15, using 
multi-object and long-slit spectroscopy observing modes.
 
 For the multi-object spectroscopic (MOS) mode we designed two masks for OB-type stellar candidates. Mask A 
 contained thirteen stars and mask B ten stars. The selection of stars considered a $K_S$ dispersion less than 2 mag 
 for the same mask to avoid large differences in the integration times and the spectral signal-to-noise ratios (S/Ns). We also 
 selected stars with reddening-free parameter $Q_{IR}$ between $-0.3$ and 0.3 (characteristic
 of OB-type stars) and $K_S$ less than 12.5 mag.
 
  The mask design also took into consideration the spectral range derived from the slit position. Slits located in the right half
 of the detector obtain spectra from 1.55 to 1.85 $\mu m$ in the $H-$band and from 2.06 to 2.40 $\mu
 m$ in the $K-$band. These spectral ranges include the \ion{He}{I} 1.70 $\mu m$, \ion{He}{I} 2.11 $\mu m$,
 \ion{He}{II} 2.57 $\mu m$, and \ion{He}{II} 1.69 $\mu m$ lines, which are required for early-type stellar spectral 
 identification and classification. 
  
  Objects for mask A were observed using the $H$ and $K$ pseudogrism, and those for mask B
 with the $H$ pseudogrism. The spectral resolution of the pseudogrisms was $\lambda/\Delta\lambda \sim
 2500$ and seeing values during observations, measured from the telluric lines in the spectra,
 were between 0.76 and 0.86 arcsec. Slits vary between 9 and 10 arcsec long and 0.85 arcsec wide.
  
  For long-slit spectroscopy we selected five bright red supergiant candidates. Stars with similar 
 magnitudes were paired to obtain spectra with similar signal-to-noise ratios. The slit was 0.75 arcsec wide and, because 
 we used the same pseudogrism as for the MOS observations, the resolution was also $\lambda/\Delta\lambda \sim 2500$.
   
  We observed with an ABBA strategy (the star is located in positions A and B in the slit, the positions then being sequentially 
 changed); with this mode, we were able to remove the sky from the spectra by subtracting spectra at 
 position A from spectra at position B, and vice versa. Flat-fielding, spectral tracing, sky subtraction, coaddition, 
 and extraction were applied using IRAF\footnote{{\sc iraf} is distributed by the National Optical Astronomy 
 Observatories, which are operated by the Association of Universities for Research in Astronomy, Inc., under 
 cooperative agreement with the National Science Foundation.} for the long-slit spectra and LIRISDR, a
 package developed specifically for LIRIS data, which uses the information from the mask design files for the 
 MOS spectra.
  
   Combining the individual spectra, we discarded cosmic rays and hot pixels that might mimic spectral lines. 
 For wavelength calibration, argon and xenon lamps were observed, both lamps 
 (continuum-subtracted) being used to 
 calibrate the $K$-band spectra and the argon lamp only for the $H$-band spectra.
  
   Three A0\,V stars were observed as telluric standards: \object{V 1431 Aql}, \object{HD 177724}, and \object{HD 167163}.
Telluric subtraction was carried out using XTELLCOR \citep{vacca03}, an IDL program that, applying a high-resolution 
synthetic model of an A0\,V star (the spectral type of our standard) over the observed telluric standards, produces 
the calibration spectrum with the telluric lines. This spectrum was then used to correct our science spectra with 
the IRAF task TELLURIC.
  
\section{Results \label{res}}

\subsection{Colour--magnitude diagram}
\label{sec:cmd}

\begin{figure*} 
\centering
\includegraphics[width=5.5in,angle=270]{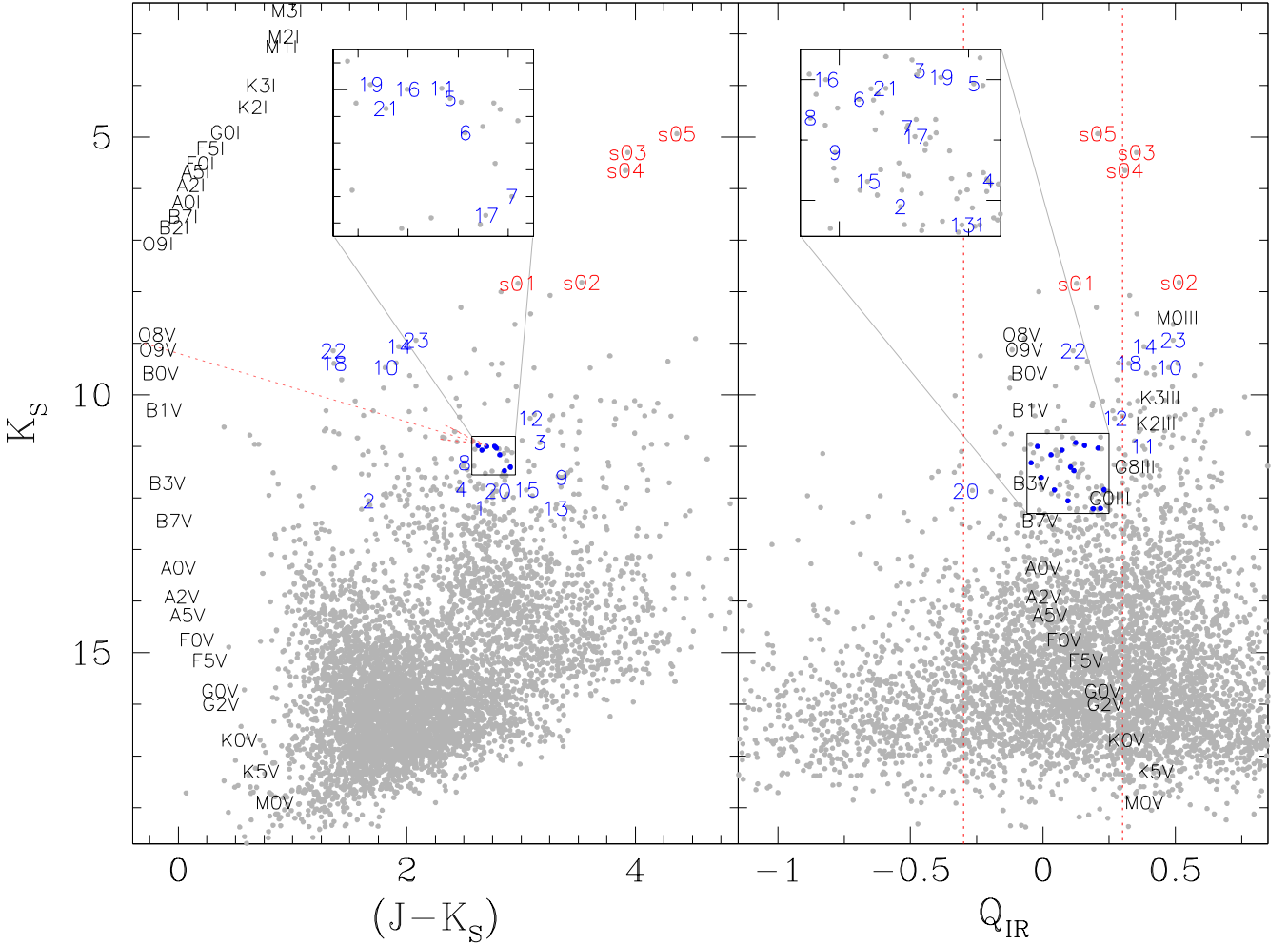} 
            \caption{Calibrated colour--magnitude (left) and free-reddening parameter--magnitude (right)
            diagrams for Masgomas-1. The red segmented arrow in the left panel shows $A_{K_S}$=3.00.
             Main sequence and supergiant sequence, shown in black, are located without reddening at the distance 
             determined in this study (3.53 kpc). In the right panel the red segmented lines show the limit 
             for the $Q_{IR}$ parameter used in the systematic search. The expected $Q_{IR}$ and $K_S$ 
             magnitude for the dwarf and giant sequences located at the distance determined in this study 
             (3.53 kpc) are also shown in black.
             In both diagrams the positions of the spectroscopically observed stars are marked with blue 
             (multi-object spectroscopy) and red numbers (long-slit spectroscopy). We marked the stars located
             in the central regions of both diagrams with blue circles and amplified the 
             regions in the upper squares, where the observed stars are marked with blue numbers. We used
              visual magnitudes and intrinsic colours from \citet{cox00} for deriving the dwarf, giant, and supergiant 
              sequences. }
       \label{msgms01_CMD}
\end{figure*}

 From the cluster's CMD we selected the OB-type candidates, using the $Q_{IR}$ selection method described
 in Section \ref{candidate_selection}. The differential extinction seems to be high for Masgomas-1, as can 
 be noted by the range of colour where the cluster candidate OB stars are distributed (see Figure 
 \ref{msgms01_CMD}). 

  In the right panel of Fig. \ref{msgms01_CMD} the $Q_{IR}$ parameter for dwarf and giant stars can be seen.
 Note that it is easy to separate both luminosity classes using the reddening-free parameter. In the same figure
 it is also noticeable that A- and early F-stars can mimic an OB star, as mentioned in Section \ref{candidate_selection};
 contamination by A- or F-type stars is expected in our set. Finally, we also find stars that satisfy the $Q_{IR}$  
 requirement, (e.g. stars with the numbers 2, 18, and 22), but their bluer colours  in the CMD might correspond to foreground stars. 
 Spectral classification for this set of stars will clarify the nature and cluster membership for these candidates.
 
 The final group of candidates corresponds to bright and red stars ($K_S\,<\,8$ mag and $(J-K_S)\,>\,3$ mag).
These bright objects are red supergiant candidates, separated from the rest of the stars by a $\gtrsim$2 mag gap
 in $K_S$. Because these stars are the brightest in the field, they would limit the spectroscopy integration time
 if they were included in the MOS, causing poor S/N spectra for the dimmer stars. To avoid this, they were
 observed using long-slit spectroscopy.
 
  The multi-object mask design also considers the positions of the candidates over 
the field of view. Because the cluster boundaries do not appear to be clearly defined in
 the LIRIS near-IR images, we included stars not only from the central region of the 
image but also from the external regions. This would help us to derive
 information about the membership of stars separated from the central part of the
 cluster. Besides this, the number and positions of slits in the mask are limited by the
spectral overlap in the dispersion axis and the different spectral range observed
as a function of the slit position in the field of view. It is necessary to ensure that the 
spectral range includes the spectral features used for the spectral classification.
   
\subsection{Spectral classification}
\label{spect_class}

\begin{figure*} 
\centering
\includegraphics[width=9.5in]{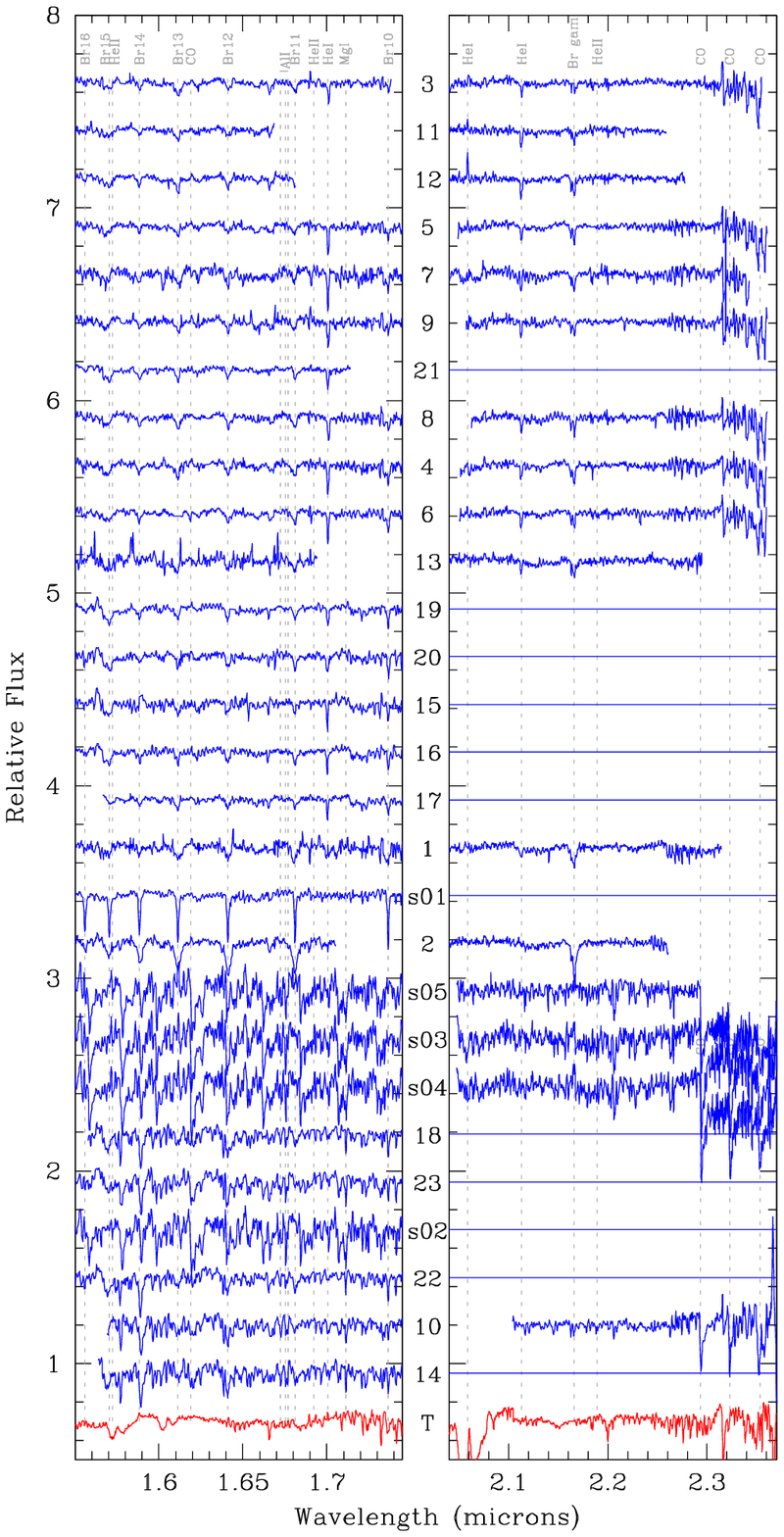} 
            \caption{Individual $H$-band (left) and $K$-band spectra (right) from mask A, mask B, and long-slit
            observations. The spectral features used for the spectral classifications are labelled in grey. 
            Spectra are arranged from early- to late-type stars. The bottom red spectrum corresponds to the 
            telluric correction.}
           \label{msgms01_masks}
\end{figure*}

We based the near-IR spectral classification of the OB stellar types on \citet{hanson96} for the $K$-band, and 
\citet{hanson98} for the $H$-band spectra. 
For later spectral types we used \citet{meyer98} and \citet{wallacehinkle97} for both bands. The 
characteristic lines for each spectral type were complemented with a visual comparison between our spectra
and other spectral catalogues, with $H$- and $K$-band spectra at similar resolutions \citep{ivanov04, ranade04, 
ranade07, hanson05}. For the assigned spectral type we assumed an error of $\pm 2$ subtypes, similar
to \citet{hanson10}, and \citet{negueruela10}. Table \ref{data_stars} contains the coordinates, 
near-infrared magnitudes, and spectral types for the spectroscopically observed stars. The positions of 
these stars in the CMD and pseudo-colour magnitude diagram are presented in Figure \ref{msgms01_CMD}. 
Spectra from masks A and B, and bright stars spectra observed with long-slit are shown in Figure \ref{msgms01_masks}. 

 One feature that required especial consideration is the spectral ghosts caused by inner reflections in the
LIRIS MOS mode. These ghosts appear as emission lines in the image and, if they coincide with the dispersion
region defined by a slit, they can mimic a spectral emission line. Because the mask is the origin of ghosts, their
positions depend on the slit positions in the mask. After a $180^{\circ}$ rotation and shifts in $\delta x = $-60
pixels and  $\delta y = $-45 pixels the slit positions coincide with the ghosts positions, making their identification
easier. 
 
 \begin{table*}
\caption{Spectroscopically observed stars. Equatorial coordinates, near-infrared magnitudes ($J$, $H$, and $K_S$), 
and spectral classification are given for all stars. For those stars with determined luminosity class are given also the
estimated extinction and distance.}
\begin{center}
\begin{tabular}{ccccccccc}
\toprule
 ID & RA (J2000) & Dec (J2000)  & $J$ & $H$ & $K_S$ & Spectral type & $A_{K_S}$ & Distance\\
      & [$\mathrm{{~~}^{h}{~~}^{m}{~~}^{s}}$] & [$\mathrm{{~~}^{\circ}~~'~~''}$] & [mag] & [mag] & [mag]  &    &       [mag]        & [kpc]  \\
 \midrule
 \multicolumn{9}{l}{Star cluster candidates:} \\
 \midrule
\vspace{0.1cm}
1   & 18 50 17.398 & +00 22 04.77 & 14.843 & 13.101 & 12.204 & B1\,V    & 1.83$^{+0.03}_{-0.02}$ & 3.69$^{+2.54}_{-1.75}$ \\ 
\vspace{0.1cm}
3   & 18 50 19.420 & +00 21 41.15 & 14.089 & 12.051 & 10.925 & O9\,V    & 2.21$^{+0.01}_{-0.03}$ & $2.95^{+0.88}_{-1.20}$ \\ 
\vspace{0.1cm}
4   & 18 50 16.434 & +00 21 31.44 & 14.316 & 12.671 & 11.839 & B0\,V     & 1.74$^{+0.02}_{-0.03}$ & $4.50^{+1.80}_{-2.14}$ \\ 
\vspace{0.1cm}
5   & 18 50 16.555 & +00 21 21.69 & 13.817 & 11.987 & 11.033 & O9.5\,V & 1.95$^{+0.01}_{-0.03}$ & $3.14^{+0.82}_{-1.08}$ \\ 
\vspace{0.1cm}
6   & 18 50 16.500 & +00 21 07.01 & 13.976 & 12.193 & 11.162 & B0\,V     & 1.96$^{+0.02}_{-0.03}$ & $2.98^{+1.19}_{-1.42}$ \\ 
\vspace{0.1cm}
7   & 18 50 17.481 & +00 20 56.26 & 14.307 & 12.438 & 11.400 & O9.5\,V & 2.03$^{+0.01}_{-0.03}$ & $3.58^{+0.94}_{-1.23}$ \\ 
\vspace{0.1cm}
8  & 18 50 15.903 & +00 20 44.31 & 13.821 & 12.261 & 11.317 & O9.5--B0\,V & 1.76$^{+0.02}_{-0.03}$ & $3.70^{+1.21}_{-1.86}$ \\
\vspace{0.1cm}
9   & 18 50 16.185 & +00 20 35.77 & 14.958 & 12.848 & 11.603 & O9.5\,V  & 2.33$^{+0.01}_{-0.03}$ & $3.43^{+0.90}_{-1.18}$ \\ 
\vspace{0.1cm}
11 & 18 50 23.945 & +00 19 57.52 & 13.762 & 11.880  & 10.995 & O9\,V & 1.95$^{+0.01}_{-0.03}$ & $3.43^{+1.02}_{-1.40}$ \\ 
\vspace{0.1cm}
12 & 18 50 23.091 & +00 19 47.64 & 13.535 & 11.496 & 10.456 & O9\,V  & 2.15$^{+0.01}_{-0.03}$ & $2.44^{+0.73}_{-0.99}$ \\ 
\vspace{0.1cm}
13 & 18 50 22.271 & +00 19 28.18 & 15.505 & 13.357 & 12.205 & B0\,V  & 2.29$^{+0.02}_{-0.03}$ & $4.15^{+1.66}_{-1.98}$ \\ 
\vspace{0.1cm}
15 & 18 50 17.303 & +00 20 59.01 & 14.889 & 12.955 & 11.843 &  B0\,V  & 2.12$^{+0.02}_{-0.03}$ & $3.79^{+1.52}_{-1.81}$ \\ 
\vspace{0.1cm}
16 & 18 50 16.837 & +00 21 12.35 & 13.697 & 12.006 & 10.999 &  B0\,V  & 1.90$^{+0.01}_{-0.01}$ & $3.51^{+1.05}_{-1.43}$ \\ 
\vspace{0.1cm}
17 & 18 50 17.036 & +00 21 46.69 & 14.326 & 12.485 & 11.471 &  B0\,V  & 2.00$^{+0.02}_{-0.03}$ & $3.39^{+1.35}_{-1.61}$ \\ 
\vspace{0.1cm}
19 & 18 50 13.554 & +00 21 03.23 & 13.606 & 11.895 & 10.981 &  B0\,V  & 1.84$^{+0.02}_{-0.03}$ & $2.90^{+1.16}_{-1.38}$ \\ 
\vspace{0.1cm}
20 & 18 50 12.329 & +00 20 46.68 & 14.829 & 13.25   & 11.96  &    B0\,V  & 2.00$^{+0.02}_{-0.03}$ & $4.23^{+1.69}_{-2.01}$ \\ 
\vspace{0.1cm}
21 & 18 50 10.573 & +00 20 25.64 & 13.727 & 12.028 & 11.071 &  O9.5\,V & 1.87$^{+0.01}_{-0.03}$ & $3.32^{+0.87}_{-1.14}$ \\ 
\vspace{0.1cm}
s01 & 18 50 17.623 & +00 21 30.65 & 10.812 &   8.893   & 7.839 & A2\,I  & 1.89$^{+0.03}_{-0.01}$ & $3.58^{+0.31}_{-0.54}$ \\ 
\vspace{0.1cm}
s03 & 18 50 15.408 & +00 20 58.07 &   9.232 &   6.625   & 5.299 & M2\,I  & 1.99$^{+0.02}_{-0.06}$ & $4.00^{+2.82}_{-0.43}$ \\ 
\vspace{0.1cm}
s04 & 18 50 15.620 & +00 21 07.46 &   9.563 &   6.984   & 5.649 & M2\,I  & 1.98$^{+0.02}_{-0.06}$ & $4.72^{+3.33}_{-0.51}$ \\ 
\vspace{0.1cm}
s05 & 18 50 14.117 & +00 22 39.56 &   9.300 &   6.477   & 4.938 & M1\,I  & 2.25$^{+0.01}_{-0.06}$ & $2.74^{+1.03}_{-0.16}$ \\ 
 \midrule
 \multicolumn{9}{l}{Fore-/background stars:} \\
 \midrule
2   & 18 50 21.558 & +00 21 52.31 & 13.720 & 12.636 & 12.054 & A0\,V    & 1.10 & 1.18 \\
10 & 18 50 14.005 & +00 20 06.94 & 11.281 &   9.966 &   9.472  &G9--K2\,III & 0.74 & 1.32 \\
14 & 18 50 20.709 & +00 21 17.59 & 10.994 &   9.638 &   9.065 & G9--K2\,III & 0.82 & 1.06 \\ 
18 & 18 50 15.788 & +00 21 46.06 & 10.752 &   9.775 &   9.391  &  G0\,III  & 0.61 & 0.81\\
22 & 18 50 10.445 & +00 21 03.80 & 10.499 &   9.603   &   9.143  & G9--K2\,III & 0.44 & 1.31 \\
23 & 18 50 08.513 & +00 22 11.42 & 11.019 &   9.528   & 8.94 &  G6\,III             & 1.01 & 0.67 \\
s02 & 18 50 28.354 & +00 21 13.44 & 11.348 &   8.935   & 7.818 & K--M I--III  & $\cdots$ & $\cdots$ \\
\bottomrule
\end{tabular}
\end{center}
\label{data_stars}
\end{table*}

  A visual inspection allowed us to separate the 28 spectra into three groups. The first one is formed
 by spectra with \ion{He}{I} 1.70 $\mu m$, \ion{He}{I} 2.11 $\mu m$, and the Brackett series; this is characteristic 
 of OB-type stars. Spectra of stars with the numbers 1, 3-9, 11-13, 15-17, and 19-21 belong to this group.
 
  The second group is formed by stars with the Brackett series as the only feature in the spectra.
Stars number 2 and s01 form this group.
 
  Finally, stars 10, 14, 18, 22, 23, s02, s03, s04, and s05 present \ion{Mg}{I} 1.58--1.71 $\mu m$ lines,
\ion{Al}{I} 1.67--1.68 $\mu m$ lines, ${}^{12}$CO $\Delta \nu=3$ (${}^{12}$CO (6,3) at 1.62 $\mu m$), and 
${}^{12}$CO $\Delta \nu=2$ bands (${}^{12}$CO (2,0) at 2.29, ${}^{12}$CO (3,1) at 2.32 and ${}^{12}$CO (4,2) at 2.35 
$\mu m$), characteristic of late type stars.
 
 The classification of stars from the first group (OB-type) is based on the Brackett series depth, and the presence 
(and depth) of \ion{He}{I} 1.70, 2.06, and 2.11 $\mu m$ lines. The absence of \ion{He}{II} lines indicates
spectral types later than O8 for all our stars.

 Stars number 3, 11, and 12 are the earliest stars from this group according to their weak Brackett series and 
the \ion{He}{I} 1.70 and 2.11 $\mu m$ lines. Only \ion{H}{I} (4-12), (4-13), and (4-14) are evident, looking similar to O8\,V
stars (v.g. \object{HD 13268}, \citealt{hanson05}), but due to the absence of \ion{He}{II} 1.69 $\mu m$ line the spectral type
cannot be earlier than O8.5\,V. The depth and shape of \ion{He}{I} 1.70 $\mu m$ and 2.11 $\mu m$ are similar
to O9.5\,V stars (for example, \object{HD 37468}, \citealt{hanson05}). Because the \ion{He}{I} 2.06 $\mu m$
emission line is not covered into the catalogue spectra, we cannot separate the spectral type for these three 
stars. Spectral type O9\,V is assigned to stars number 3, 11, and 12.

 Stars number 5, 7, 8, 9, and 21 display the deepest \ion{He}{I} 1.70 $\mu m$ of the spectral set. Inspection of
the sky spectra from their slits discarded the contamination by nebular \ion{He}{I} emission, which could have produced
an excess of sky correction and thus extra absorption. The depths of this \ion{He}{I}
line, together with Brackett lines \ion{H}{I} (4-11) and \ion{H}{I} (4-10), resemble the spectrum of O9.5\,V stars (for example,
\object{HD 37468}, \citealt{hanson05}). For stars number 5, 7, 9, and 21 we assigned spectral type
O9.5\,V. In the case of star number 8 the Brackett series is slightly deeper, similar to \object{HD 149438} and \object{HD 36822} 
(B0.2\,V stars, \citealt{hanson05}), hence the spectral type for this star is between O9.5 and B0\,V.
 
 For stars number 13, 19, and 20, the Brackett series looks deeper than for stars number 11 and 12. The depths of \ion{He}{I} 2.11 $\mu m$
and \ion{H}{I} (4-7) are similar to spectral B0\,V types (for example, \object{HD 6165} \citealt{wallacehinkle97}).
Spectral type B0\,V was therefore adopted for these stars.

  Spectra from stars number 1, 4, and 6 show clear Brackett series until \ion{H}{I} (4-15). The \ion{H}{I} (4-7) line is deeper than
\ion{He}{I} 2.11 $\mu m$, indicating dwarf luminosity class (e.g. B1\,V star \object{HD 191639} and B1\,V \object{HD 31726};
 \citealt{hanson96}). The similarity of the Brackett series and \ion{He}{I} lines to the mentioned
stars implies a B1\,V spectral type for star number 1. For stars number 4 and 6, the depth of \ion{He}{I} 1.70 $\mu m$ and
the similar \ion{H}{I} (4-10) line to stars number 13, 19, and 20 imply an earlier spectral type (B0\,V).  

 Finally, three stars from this group (stars number 15, 16, and 17) exhibit narrow \ion{He}{I} 1.70 $\mu m$ that would
indicate luminosity class III. Unfortunately, the lack of $K$-band spectra - and the associated \ion{He}{I} 2.11 $\mu m$ line -
does not allow a clear differentiation between luminosity classes V and III. Nonetheless, the individual distance estimates
for these three stars under the assumption of a luminosity class III are higher (5.80, 5.16 and 6.12 kpc for stars number 15, 16, 
and 17) than the calculated for luminosity class V, which agrees with the rest of the stars. Because 
for both luminosity classes the estimated individual extinctions are the same, we consider it to be highly unlikely that these three 
stars belong to another cluster located at $\sim5.5$ kpc, in the same line of sight as Masgomas-1, without additional
extinction. These arguments, together with the central position of the stars in the cluster's field of view favour luminosity class V 
for these three stars. The clear Brackett series between \ion{H}{I} (4-15) and \ion{H}{I} (4-10) implies a spectral type B0\,V
 
 The spectrum of star number 2 presents neither \ion{He}{I} nor \ion{He}{II} lines. The \ion{H}{I} (4--7) line indicates a star later than
B8\,V (\object{HD 169990}, \citealp{hanson05}), and fits an A0\,V star (for example, \object{HR 5793}, \citealp{meyer98}).
Later spectral types show a deeper line. The Brackett series fits an A0\,V star from \ion{H}{I} (4-18) to \ion{H}{I} (4-11)
 (e.g. \object{HD 122945}, previously observed with LIRIS, and \object{HR 7001}, \citealp{meyer98}). The blue colour
 for this star compared with the other observed stars and the A0\,V spectral type indicate that star number 2
 is a foreground star.
   
   As mentioned previously, stars number 10, 14, 18, 22, and 23 show lines characteristic of late-type stars 
(e.g. \ion{Mg}{I} 1.58--1.71 $\mu m$, \ion{Al}{I} 1.67--1.68 $\mu m$, and ${}^{12}$CO $\Delta \nu=3$ bands). 
The ${}^{12}$CO $\Delta \nu=2$ bands of star number 10 are similar in depth to an early K III star. The \ion{Mg}{I} 
1.58--1.71--2.28 $\mu m$ and ${}^{12}$CO $\Delta \nu=3$ bands are characteristic of G9--K2 III stars (e.g. 
\object{HR 7328}, \object{HR 5340} or \object{HR 7806}, \citealt{meyer98}). Star number 10 is classified as a  
foreground G9--K2 III source.

 The $H$-band spectra of stars number 14 and 22 are similar to the spectrum of star number 10. A better spectral classification
than obtained for star number 10 without the $K$-band spectra is not possible. Hence, for stars number 14 and 22 we assigned
the same spectral type as for star number 10.
 
 The spectra of stars number 18 and 23 look similar to the spectrum of star number 10 but with minor differences.
Star number 18 shows a smaller ${}^{12}$CO $\Delta \nu=2$ band and \ion{Mg}{I} 1.58--1.71 $\mu m$ lines. The 
$H$-band spectrum is similar to that of a G0\,III star (for example, \object{HR 4883}, \citealt{meyer98}), 
which is the spectral type assigned to star number 18. The spectrum of star number 23 has deeper \ion{Mg}{I} 1.58--1.71 
$\mu m$ lines than star number 18. Because its spectrum looks similar to that of star \object{HR 4716} (G6\,III star, \citealt{meyer98}),
we assigned  a G6\,III spectral type to this star.
	
 Long-slit spectra for the five bright stars (s01, s02, s03, s04, and s05) show the characteristics of giant/supergiant stars.
 For example, the spectrum of s01 is dominated by a narrow Brackett series, typical of luminosity class I \citep{meyer98}.   
 The absence of \ion{He}{I} 1.70 $\mu m$ discounts a late B-type star, and the depth of the hydrogen lines is similar to that for
 an A I star. The depths of the \ion{Ne}{II} line at 1.77 $\mu m$ and the \ion{H}{I} (4--10), (4--11), (4--12), and (4--13) lines 
 indicate an A2 spectral type for s01 (like star \object{HR 7924}, \citealt{meyer98}).
      
 The spectra of s02, s03, s04 and s05 show strong ${}^{12}$CO $\Delta \nu=3$ and ${}^{12}$CO $\Delta \nu=2$ (for s03, 
s04 and s05) bands associated with late K--M giant and supergiant stars. For s03, s04 and s05, the equivalent width (EW) 
of ${}^{12}$CO (2,0) corresponds to luminosity class I. For them we measured $\mathrm{{EW}_{s03}}=31.37 \AA$, 
$\mathrm{{EW}_{s04}}=31.47 \AA$ and $\mathrm{{EW}_{s05}}=27.89 \AA$, in the region between 2.294 and 2.304 ${\mu m}$. 
For early M-stars, these values lie in the supergiant zone in the relation given by \citet{davies07}, but differences in the spectral 
lines and molecular band depth for the early M-star subclasses are hard to find at our spectral resolution. For s03 and s04 the 
depth of the ${}^{12}$CO $\Delta \nu=3$ bands is more similar to an M2 (e.g. \object{HD 14479}; \citealt{meyer98}) than
to an M1 star (e.g. \object{HD 339034}; \citealt{meyer98}) or to an M4-5 source (e.g. \object{HR 7009}; \citealt{meyer98}). 
The depth of s05 is indicative of an earlier spectral subtype (M1\,I). 
  
  For star s02 the luminosity class is not straightforward to define because of the lack of 
  $K$-band spectroscopy. The softer continuum (compared with s03 and s04) and the distance from the cluster centre
  ($\sim3.2$ arcmin) support a fore-/background giant star classification for s02, but we cannot clearly decide
   between luminosity classes III or I  for this object.
  
\subsection{Proper motion}
 To measure OB-type and supergiant stars proper motions, we followed a similar procedure as described
 by \citet{penaramirez11}. We used the 2MASS image as the first epoch and the LIRIS image as the second epoch, with a time
baseline of 10.88 years. The spatial transformation between both epochs image coordinate systems was performed with 
370 resolved stars well distributed over the whole LIRIS image, with $K_S<14$. For stellar positions we used 
the centroids given in the 2MASS catalogue and the coordinates derived from LIRIS photometry. Objects classified as early-type
and supergiants were excluded from the set of calibration stars to avoid the inclusion of cluster stars in the transformation calculation. 
The dispersion of the transformation is 0.25 pix for $x$ and 0.30 pix for $y$. Because these values are higher than the errors in
the object's centroid determination, we adopted them as mean proper motion errors.

\begin{figure}
\centering
\includegraphics[width=3.5in]{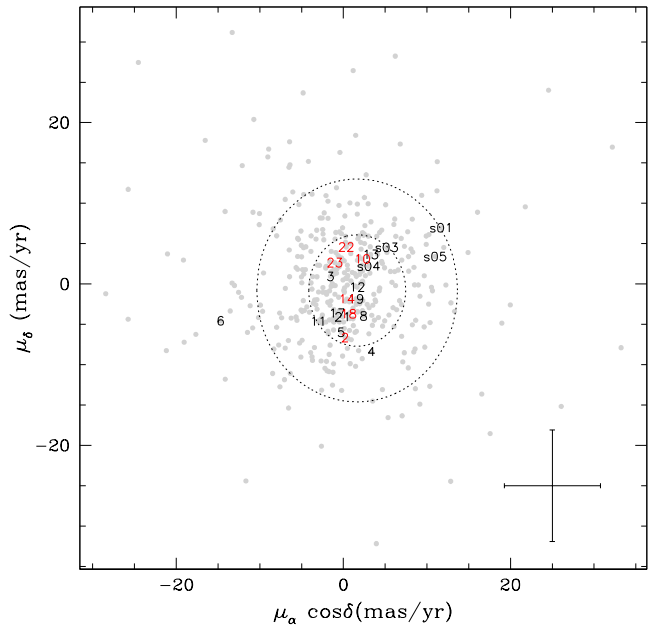}
            \caption{ Proper motions for OB-type and supergiant stars of Masgomas-1. Grey circles indicate the 
            proper motions for stars without spectral follow-up (unidentified cluster members, foreground and 
            background stars), and black numbers show the proper motion for cluster massive stars, classified spectroscopically. 
            Red numbers indicate the proper motion for foreground stars, derived from the spectroscopic observations.
            In the bottom right corner of the figure we show the mean error for the proper motions ($\Delta (\mu_{\alpha} 
            cos\delta)=5\,\farcs75$ and $\Delta (\mu_{\delta})=6\,\farcs91$). Blue ellipses represent 1$\sigma$ and 2$\sigma$ 
            proper motion uncertainties.}
       \label{prop_motion}
    \end{figure}

 After obtaining the parameters for the transformation, we calculated the pixel shifts for the four supergiants 
(stars s01, s03, s04, and s05) and the OB-type stars (stars number 3-6, 8, 11-13, 17, and 19), 
OB-type stars without 2MASS photometry were not included in the proper motion determination.
  To finally convert the pixel shifts into proper motions, we divided by the time baseline 
between both epochs (i.e. 10.88 years), and multiplied by the pixel scale of LIRIS image. 
The obtained proper motions are presented in Figure \ref{prop_motion}. 


In Figure \ref{prop_motion} we present in grey dots the proper motions for the references stars.  This group of stars could 
include unidentified cluster members, foreground and background stars. Proper motions for the identified cluster stars and
six foreground stars are labelled with the same identification numbers as used in Section \ref{spect_class}. In the figure we can 
see that most of the LIRIS field of view stars are within the 2$\sigma$ uncertainty ellipse. No proper motion difference
between cluster and fore-/background stars can be detected with our resolution.

 \section{Discussion \label{discussion}}
  \subsection{Distance estimate}
  \label{distance_estimate}

Using the estimated spectral types, we derived  individual distances for the cluster stars. Assuming the 
absolute visual magnitudes from \citet{cox00}, intrinsic infrared colours from \citet{tokunaga00}, and 
the \citet{rieke89} extinction law with $R=3.09$ \citep{rieke85}, the extinction for the $K_S$ band may be 
expressed as

\begin{equation}
A_{K_S} = \frac{E_{J-K_S}}{1.514} = \frac{E_{H-K_S}}{0.561} .
\end{equation}

 Extinction ranges from 1.10 to 2.33 mag for $A_{K_S}$, or $A_V$ between 10.15 and 21.49 mag.
In Table \ref{data_stars} we present the values of extinction and individual distances
 for the OB-type stars and the supergiant stars. The mean of these 
individual distances is $3.53^{+1.55}_{-1.40}$ kpc, our estimate for the distance to Masgomas-1. For comparison
we estimated the individual distances using the \citet{indebetouw05} extinction law, obtaining lower
values for them. The mean distance estimated using this extinction law is 
$3.50^{+1.55}_{-1.40}$ kpc, which is consistent with the estimate obtained using the \citet{rieke89} 
extinction law to within the errors. Late-type giant stars were excluded from the distance estimate 
because their individual distance estimates imply that they are not part of the cluster and probably
 belong to the disc population.
 
 Even if Masgomas-1 is located in the same direction as the base of the Scutum--Centaurus Arm
our estimate positions the cluster closer than the red supergiant clusters (RSGC). Distances for RSGC 1, 
2, 3 and 4 (6.60, 5.83, 6, and 6.6 kpc respectively; \citealp{clark09a,negueruela10}) indicate that they 
belong to the intersection of the tip of the Galactic bar and the base of the Scutum--Centaurus Arm.
 Our estimate of 3.53 kpc and the galactic coordinates for Masgomas-1 place it in the Scutum--Centaurus
arm itself but without any evidence relating it with the star-forming region located in the intersection
between the end of the Galactic bar and the base of the Scutum--Centaurus arm.
 

\subsection{Mass and age estimate}

We estimated the total mass of Masgomas-1 by integrating the initial mass functions, fitted to the cluster
mass function. This was performed separately for a Salpeter and a Kroupa initial mass function (IMF).
The Salpeter IMF \citep{salpeter55} was fitted to the cluster's massive population: the O-type dwarf and the supergiant stars. We used
this IMF for the first estimate to compare it with the values given by \object{RSGC 3} and \object{Cl Alicante 8}.

 The O-type dwarf population is formed by eight stars with spectral type between O9--O9.5\,V and mass
between 15 and $18 M_{\odot}$ \citep{martins05}. The integration of a Salpeter function indicates that
a $11000 M_{\odot}$ initial mass is expected for the cluster to have a population of eight O-type dwarfs.
This number could be underestimated, because we only included those stars in the population that were
observed spectroscopically and with the subsequent spectral classification.
 
 For the supergiant population, formed by stars s01, s03, s04 and s05, the same method indicates that 
a cluster of $8000-9000 M_{\odot}$ is required to host a population of four supergiant stars, with initial 
mass of $\sim20 M_{\odot}$, determined using evolutionary tracks from \citet{marigo08}.

 In both populations we integrated the Salpeter IMF from log $(M) = -1.0$ dex to 1.3 dex. 
The first limit corresponds to the stellar lower limit ($\sim 0.1 M_{\odot}$) and the second one 
corresponds to the most massive star detected in our work for Masgomas-1. Our estimate 
for the total mass of Masgomas-1 is between 8000 and 11000 ${M}_{\odot}$.
 
 This estimate agrees with the initial total mass obtained for other clusters with a red supergiant
population. For \object{RSGC 3}, \citet{clark09a} estimated a total mass of $(2-4)\cdot 10^4 {M}_{\odot}$, 
and for \object{Cl Alicante 8} \citep{negueruela10} the total initial mass is estimated to be $2\cdot 10^4 {M}_{\odot}$.
Both clusters contains eight supergiants, therefore a higher initial total mass is expected. Our total mass estimate
also agrees with the simulations presented by \citet{clark09b}, where three red supergiants are
expected for clusters with a total mass over $10^4{M}_{\odot}$.

 The second estimate of the cluster total mass was obtained using the cluster initial mass function,
ranging from $\sim3{M}_{\odot}$ to $\sim30{M}_{\odot}$. The mass
function was derived from the luminosity function, corrected for the field stellar contribution using a control field.  
We fitted a Kroupa IMF \citep{kroupa01} to the mass histogram and integrated within the 
same limits used for our first estimate (i.e. from log $(M) = -1.0$ dex to 1.3 dex). Because 
we had no LIRIS photometry for a control field, we used UKIDSS\footnote{The UKIDSS project is 
defined in \citet{lawrence07}. UKIDSS uses the UKIRT Wide Field Camera (WFCAM; \citealp{casali07}). The 
photometric system is described in \citet{hewett06}, and the calibration is described in \citet{hodgkin09}. The 
pipeline processing and science archive are described in \citet{hambly08}.} (for $K>9.3$) and 2MASS (for $K_S<9.3$) 
photometry for both Masgomas-1 and control fields. The circular area for the control field is centred on 
${\alpha}_{2000}=18^{\mathrm {h}}49^{\mathrm {m}}47^{\mathrm {s}}$, ${\delta}_{2000}=+00\degr13\arcmin36\arcsec$,
and it has a radius of 3$\arcmin$. The same radius was used to obtain the photometry for the Masgomas-1 field.

 To correct the difference between both photometric systems, we used the transformation
equations given by \citet{carpenter01}. For the UKIDSS photometry we used data for $J<18.1$, 
$H<16.8$, and $K<16.1$, to ensure completeness close to 1.0 for the data. These limiting magnitudes
were determined from the star count histograms for each filter, the magnitude where the star counts start to decrease
was defined as the limiting magnitude.

 To obtain the luminosity and mass functions for both fields, we followed the same procedure as
described by \citet{ramirezalegria11}. First, we projected every star following the reddening 
vector to the dwarf star sequence, located at our estimated distance, and defined by the magnitudes 
and colours of \citet{cox00}. This sequence is expressed analytically by two lines, one from O9\,V 
to A0\,V and the second from A0\,V to G2\,V. The cut in G2\,V arises from the selected limiting
magnitude for the UKIDSS photometry.

 Once the stars in the Masgomas-1 and the control field CMDs were projected following the 
reddening vector to the line-fitted main sequence, we derived the luminosity functions.
We converted the $K$ magnitude to stellar mass using the values given by \citet{cox00}, 
obtaining  the mass functions. For magnitudes that were in between values from the
catalogue, we interpolated between the two closest values.

\begin{figure}
\centering
\includegraphics[width=3.5in]{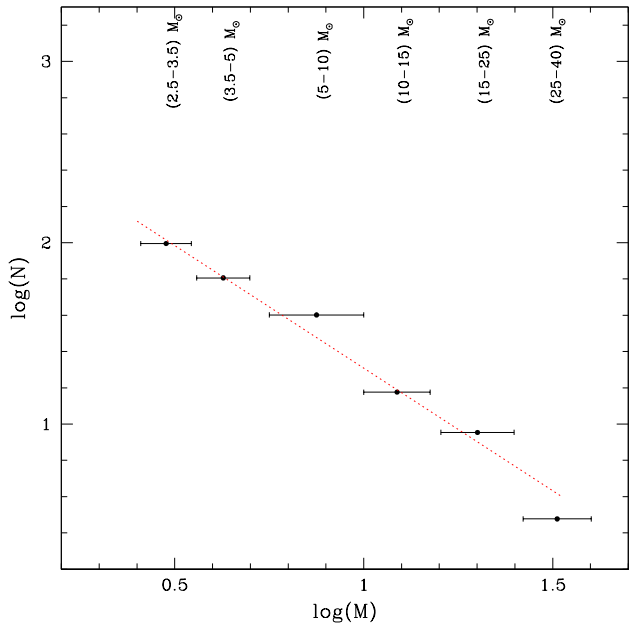}
            \caption{Mass function for Masgomas-1. The points indicate the central position of the mass range 
            indicated above. The segmented red line shows the Kroupa IMF fitted to the data.}
       \label{histograms}
    \end{figure}

After subtracting both mass functions (i.e. Masgomas-1 field minus control field), we obtained 
the cluster initial mass function (shown in Fig. \ref{histograms}). We fitted a multiple segment 
Kroupa IMF\footnote{As mentioned by \citet{clark09b}, 
an estimate of the total mass cluster by integrating a Salpeter IMF overestimates the low-mass stars contribution,
 resulting in a difference of $\sim30 \%$ between the two total mass estimates.}, and integrated it in the range 0.10 to 20 ${M}_{\odot}$, 
obtaining a total mass for Masgomas-1 of $(1.94\pm0.28)\cdot10^4 {M}_{\odot}$. This value is higher
than estimated by fitting a Salpeter IMF only to the massive cluster population, but it considers
a wider range of mass for the cluster population, and confirms the massive nature of the cluster.

 For the age estimate we derived information based on the earliest main-sequence star or the 
presence of red supergiant stars. In the first case, the O9\,V star indicates an age upper limit 
of 10 Myr, which is the time spent by an O9 star in the main sequence, depending on the initial
rotational velocity \citep{brott11}.

 On the other hand, the M-type red supergiants position in the CMD could be described with an isochrone, but
caution has to be taken because the supergiant's intrinsic colours are not well constrained. As mentioned by
\citet{negueruela10}, the extinction correction for supergiants can be made incorrectly due to colour terms or 
the structure of their atmospheres. Because of that, we prefer not to estimate the cluster age by fitting an 
isochrone to the predicted intrinsic magnitudes and colours of the red supergiants.

 According to \citet{davies07}, red supergiants are expected in massive clusters after $\sim$6 Myr., which is a lower limit
that is compatible with the upper limit set by the O9\,V star. In their Fig. 2, we can see that a $2\cdot10^4{M}_{\odot}$
cluster of $\sim6.5$ Myr should contain 3 or 4 red supergiants. For less massive clusters the needed age for
that number of red supergiants would be higher, but hardly higher than the upper limit set by the earliest main
sequence star in Masgomas-1. For example Fig. 3 from \citet{clark09b} shows that a $10^4{M}_{\odot}$
 cluster would present three red supergiants after $\sim10$ Myr, still an upper limit consistent with the one set by the presence
 of an O9\,V cluster star.

With these arguments, we estimate an age for Masgomas-1 between 8 to 10 Myr.


\section{Conclusions \label{conclusions}}

As part of our systematic search and physical characterization of massive stellar clusters,
we completed a spectrophotometric study of Masgomas-1, the first cluster
discovered by our MASGOMAS project. This study was completed using photometric and
spectroscopic LIRIS data obtained at the WHT.

 With the new near-infrared data and the reddening-free $Q_{IR}$ parameter, we 
selected 23 OB-type candidates for $H$ and $K$ spectroscopy. Five bright red stars
from the Masgomas-1 field of view were selected as red supergiant candidates and
for $H$ and $K$ spectroscopy. These 28 stars were spectrally classified and individual
spectrophotometric distances were estimated for the OB-type and supergiant stars.
The individual distances and the associated individual extinctions are consistent 
with a common distance for all objects. For all supergiants (one yellow and three
red) and ten of the OB-type stars, we measured proper motions. The resolution of the 
proper motion measures did not allow us to find differences between the individual distances
of these stars.
 
  Using the individual distance estimate, we obtained a distance to Masgomas-1 of
$ 3.53^{+1.55}_{-1.40}$ kpc, placing Masgomas-1 in the Scutum--Centaurus arm, but far from 
the arm base.
 
  For the cluster total mass we obtained a lower limit, adjusting a Salpeter function to the
most massive stars in our work. Integrating this function between 20 and 0.1 $M_{\odot}$,
we obtained a total mass of $(0.8-1.1)\cdot10^4 {M}_{\odot}$. We repeated this
procedure using the initial cluster mass function fitted with a Kroupa function. Integration
of this function in the same mass range confirms the massive nature of Masgomas-1 (i.e. over 
$10^4 {M}_{\odot}$), and gives a higher estimate for the cluster's total mass: $(1.94\pm0.28)\cdot10^4 {M}_{\odot}$. 

The cluster age could be limited in two ways: an upper limit given by the earliest star in the cluster's 
main sequence of 10 Myr, and a lower limit given by the presence of the M-type supergiants, estimated 
between 6.5 and 10 Myr. The latest limit varies according to the total initial mass of the cluster, which remains
compatible with the upper limit of 10 Myr set by the presence of the O9\,V star.

 
\begin{acknowledgements}

S.R.A. was supported by the investigation project Massive Stars in Galactic Obscured Massive Clusters.
Part of this work was supported by the Science and Technology Ministry of the Kingdom of Spain (grants 
AYA2008-06166-C03-01, and AYA2010-21697-C05-04), the Gobierno de Canarias (PID2010119),
 and the Fundaci\'on Agencia Aragonesa para la Investigaci\'on y Desarrollo 
(ARAID). S.R.A., A.H., and A.M-F. are members of the Consolider-Ingenio 2010 Programme (CSD2006-00070).

 The William Herschel Telescope is operated on the island of La Palma by the Isaac Newton Group in the 
 Spanish Observatorio del Roque de los Muchachos of the Instituto de Astrof\'isica de Canarias. This publication 
 makes use of data products from the Two Micron All Sky Survey, which is a joint project of the University of 
 Massachusetts and the Infrared Processing and Analysis Center/California Institute of Technology, funded by the 
 National Aeronautics and Space Administration and the National Science Foundation.

\end{acknowledgements}

\bibpunct{(}{)}{;}{a}{}{,} 
\bibliographystyle{aa} 
\bibliography{biblio}

\begin{thebibliography}{42}
\expandafter\ifx\csname natexlab\endcsname\relax\def\natexlab#1{#1}\fi

\bibitem[{{Benjamin} {et~al.}(2003){Benjamin}, {Churchwell}, {Babler}, {Bania},
  {Clemens}, {Cohen}, {Dickey}, {Indebetouw}, {Jackson}, {Kobulnicky},
  {Lazarian}, {Marston}, {Mathis}, {Meade}, {Seager}, {Stolovy}, {Watson},
  {Whitney}, {Wolff}, \& {Wolfire}}]{benjamin03}
{Benjamin}, R.~A., {Churchwell}, E., {Babler}, B.~L., {et~al.} 2003, \pasp,
  115, 953

\bibitem[{{Brott} {et~al.}(2011){Brott}, {de Mink}, {Cantiello}, {Langer}, {de
  Koter}, {Evans}, {Hunter}, {Trundle}, \& {Vink}}]{brott11}
{Brott}, I., {de Mink}, S.~E., {Cantiello}, M., {et~al.} 2011, \aap, 530, A115

\bibitem[{{Carpenter}(2001)}]{carpenter01}
{Carpenter}, J.~M. 2001, \aj, 121, 2851

\bibitem[{{Casali} {et~al.}(2007){Casali}, {Adamson}, {Alves de Oliveira},
  {Almaini}, {Burch}, {Chuter}, {Elliot}, {Folger}, {Foucaud}, {Hambly},
  {Hastie}, {Henry}, {Hirst}, {Irwin}, {Ives}, {Lawrence}, {Laidlaw}, {Lee},
  {Lewis}, {Lunney}, {McLay}, {Montgomery}, {Pickup}, {Read}, {Rees}, {Robson},
  {Sekiguchi}, {Vick}, {Warren}, \& {Woodward}}]{casali07}
{Casali}, M., {Adamson}, A., {Alves de Oliveira}, C., {et~al.} 2007, \aap, 467,
  777

\bibitem[{Clark {et~al.}(2009a)Clark, Negueruela, Davies, Larionov, Ritchie,
  Figer, Messineo, Crowther, \& Arkharov}]{clark09a}
Clark, J.~S., Negueruela, I., Davies, B., {et~al.} 2009a, \aap, 498, 109

\bibitem[{{Clark} {et~al.}(2009b){Clark}, {Davies}, {Najarro}, {MacKenty},
  {Crowther}, {Messineo}, \& {Thompson}}]{clark09b}
{Clark}, J.~S., {Davies}, B., {Najarro}, F., {et~al.} 2009b, \aap, 504, 429

\bibitem[{{Comer{\'o}n} \& {Pasquali}(2005)}]{comeron05}
{Comer{\'o}n}, F. \& {Pasquali}, A. 2005, \aap, 430, 541

\bibitem[{{Cox}(2000)}]{cox00}
{Cox}, A.~N. 2000, Allen's Astrophysical Quantities, 4th edition (Springer, New York)

\bibitem[{Davies {et~al.}(2007)Davies, Figer, Kudritzki, MacKenty, Najarro, \&
  Herrero}]{davies07}
Davies, B., Figer, D.~F., Kudritzki, R.-P., {et~al.} 2007, \apj, 671, 781

\bibitem[{{Davies} {et~al.}(2008){Davies}, {Figer}, {Law}, {Kudritzki},
  {Najarro}, {Herrero}, \& {MacKenty}}]{davies08}
{Davies}, B., {Figer}, D.~F., {Law}, C.~J., {et~al.} 2008, \apj, 676, 1016

\bibitem[{{Eikenberry} {et~al.}(2006){Eikenberry}, {Elston}, {Raines},
  {Julian}, {Hanna}, {Hon}, {Julian}, {Bandyopadhyay}, {Bennett}, {Bessoff},
  {Branch}, {Corley}, {Eriksen}, {Frommeyer}, {Gonzalez}, {Herlevich},
  {Marin-Franch}, {Marti}, {Murphey}, {Rashkin}, {Warner}, {Leckie},
  {Gardhouse}, {Fletcher}, {Dunn}, {Wooff}, \& {Hardy}}]{eikenberry06}
 {Eikenberry}, S., {Elston}, R., {Raines}, S.~N., {et~al.} 2006, in 
Ground-based and Airborne Instrumentation for Astronomy, ed.
{I.~S.~McLean, \& M.~Ian} (SPIE, Washington), Proceedings of the SPIE, 6269, 626917

\bibitem[{{Figer} {et~al.}(2006){Figer}, {MacKenty}, {Robberto}, {Smith},
  {Najarro}, {Kudritzki}, \& {Herrero}}]{figer06}
{Figer}, D.~F., {MacKenty}, J.~W., {Robberto}, M., {et~al.} 2006, \apj, 643,
  1166

\bibitem[{{Hambly} {et~al.}(2008){Hambly}, {Collins}, {Cross}, {Mann}, {Read},
  {Sutorius}, {Bond}, {Bryant}, {Emerson}, {Lawrence}, {Rimoldini}, {Stewart},
  {Williams}, {Adamson}, {Hirst}, {Dye}, \& {Warren}}]{hambly08}
{Hambly}, N.~C., {Collins}, R.~S., {Cross}, N.~J.~G., {et~al.} 2008, \mnras,
  384, 637

\bibitem[{Hanson {et~al.}(1996)Hanson, Conti, \& Rieke}]{hanson96}
Hanson, M.~M., Conti, P.~S., \& Rieke, M.~J. 1996, \apjs, 107, 281

\bibitem[{{Hanson} {et~al.}(2005){Hanson}, {Kudritzki}, {Kenworthy}, {Puls}, \&
  {Tokunaga}}]{hanson05}
{Hanson}, M.~M., {Kudritzki}, R., {Kenworthy}, M.~A., {Puls}, J., \&
  {Tokunaga}, A.~T. 2005, \apjs, 161, 154

\bibitem[{{Hanson} {et~al.}(2010){Hanson}, {Kurtev}, {Borissova}, {Georgiev},
  {Ivanov}, {Hillier}, \& {Minniti}}]{hanson10}
{Hanson}, M.~M., {Kurtev}, R., {Borissova}, J., {et~al.} 2010, \aap, 516, A35

\bibitem[{{Hanson} \& {Popescu}(2008)}]{hansonpopescu08}
 {Hanson}, M.~M. \& {Popescu}, B. 2008, in Massive Stars as Cosmic Engines, ed. 
{F.~Bresolin, P.~A.~Crowther, \& J.~Puls} (Cambridge Univ. Press, Cambridge), IAU Symposium, 250, 307

\bibitem[{Hanson {et~al.}(1998)Hanson, Rieke, \& Luhman}]{hanson98}
Hanson, M.~M., Rieke, G.~H., \& Luhman, K.~L. 1998, \aj, 116, 1915

\bibitem[{{Hewett} {et~al.}(2006){Hewett}, {Warren}, {Leggett}, \&
  {Hodgkin}}]{hewett06}
{Hewett}, P.~C., {Warren}, S.~J., {Leggett}, S.~K., \& {Hodgkin}, S.~T. 2006,
  \mnras, 367, 454

\bibitem[{{Hodgkin} {et~al.}(2009){Hodgkin}, {Irwin}, {Hewett}, \&
  {Warren}}]{hodgkin09}
{Hodgkin}, S.~T., {Irwin}, M.~J., {Hewett}, P.~C., \& {Warren}, S.~J. 2009,
  \mnras, 394, 675

\bibitem[{{Indebetouw} {et~al.}(2005){Indebetouw}, {Mathis}, {Babler}, {Meade},
  {Watson}, {Whitney}, {Wolff}, {Wolfire}, {Cohen}, {Bania}, {Benjamin},
  {Clemens}, {Dickey}, {Jackson}, {Kobulnicky}, {Marston}, {Mercer},
  {Stauffer}, {Stolovy}, \& {Churchwell}}]{indebetouw05}
{Indebetouw}, R., {Mathis}, J.~S., {Babler}, B.~L., {et~al.} 2005, \apj, 619,
  931

\bibitem[{{Ivanov} {et~al.}(2004){Ivanov}, {Rieke}, {Engelbracht},
  {Alonso-Herrero}, {Rieke}, \& {Luhman}}]{ivanov04}
{Ivanov}, V.~D., {Rieke}, M.~J., {Engelbracht}, C.~W., {et~al.} 2004, \apjs,
  151, 387

\bibitem[{{Kroupa}(2001)}]{kroupa01}
{Kroupa}, P. 2001, \mnras, 322, 231

\bibitem[{{Lawrence} {et~al.}(2007){Lawrence}, {Warren}, {Almaini}, {Edge},
  {Hambly}, {Jameson}, {Lucas}, {Casali}, {Adamson}, {Dye}, {Emerson},
  {Foucaud}, {Hewett}, {Hirst}, {Hodgkin}, {Irwin}, {Lodieu}, {McMahon},
  {Simpson}, {Smail}, {Mortlock}, \& {Folger}}]{lawrence07}
{Lawrence}, A., {Warren}, S.~J., {Almaini}, O., {et~al.} 2007, \mnras, 379,
  1599

\bibitem[{{Marigo} {et~al.}(2008){Marigo}, {Girardi}, {Bressan}, {Groenewegen},
  {Silva}, \& {Granato}}]{marigo08}
{Marigo}, P., {Girardi}, L., {Bressan}, A., {et~al.} 2008, \aap, 482, 883

\bibitem[{Mar{\'\i}n-Franch {et~al.}(2009)Mar{\'\i}n-Franch, Herrero, Lenorzer,
  Najarro, Ramirez, Font-Ribera, \& Figer}]{marin09}
Mar{\'\i}n-Franch, A., Herrero, A., Lenorzer, A., {et~al.} 2009, \aap, 502, 559

\bibitem[{Martins {et~al.}(2005)Martins, Schaerer, \& Hillier}]{martins05}
Martins, F., Schaerer, D., \& Hillier, D.~J. 2005, \aap, 436, 1049

\bibitem[{Meyer {et~al.}(1998)Meyer, Edwards, Hinkle, \& Strom}]{meyer98}
Meyer, M.~R., Edwards, S., Hinkle, K.~H., \& Strom, S.~E. 1998, \apj, 508, 397

\bibitem[{Negueruela {et~al.}(2010)Negueruela, Gonz{\'a}lez-Fern{\'a}ndez,
  Marco, Clark, \& Mart{\'\i}nez-N{\'u}{\~n}ez}]{negueruela10}
Negueruela, I., Gonz{\'a}lez-Fern{\'a}ndez, C., Marco, A., Clark, J.~S., \&
  Mart{\'\i}nez-N{\'u}{\~n}ez, S. 2010, \aap, 513, 74

\bibitem[{{Negueruela} \& {Schurch}(2007)}]{negueruela07}
{Negueruela}, I. \& {Schurch}, M.~P.~E. 2007, \aap, 461, 631

\bibitem[{{Pe{\~n}a Ram{\'{\i}}rez} {et~al.}(2011){Pe{\~n}a Ram{\'{\i}}rez},
  {Zapatero Osorio}, {B{\'e}jar}, {Rebolo}, \& {Bihain}}]{penaramirez11}
{Pe{\~n}a Ram{\'{\i}}rez}, K., {Zapatero Osorio}, M.~R., {B{\'e}jar}, V.~J.~S.,
  {Rebolo}, R., \& {Bihain}, G. 2011, \aap, 532, A42

\bibitem[{{Ram{\'{\i}}rez Alegr{\'{\i}}a} {et~al.}(2011){Ram{\'{\i}}rez
  Alegr{\'{\i}}a}, {Herrero}, {Mar{\'{\i}}n-Franch}, {Puga}, {Najarro}, {Acosta
  Pulido}, {Hidalgo}, \& {Sim{\'o}n-D{\'{\i}}az}}]{ramirezalegria11}
{Ram{\'{\i}}rez Alegr{\'{\i}}a}, S., {Herrero}, A., {Mar{\'{\i}}n-Franch}, A.,
  {et~al.} 2011, \aap, 535, A8

\bibitem[{{Ranada} {et~al.}(2007){Ranada}, {Singh}, {Gupta}, \&
  {Ashok}}]{ranade07}
{Ranada}, A.~C., {Singh}, H.~P., {Gupta}, R., \& {Ashok}, N.~M. 2007, Bulletin
  of the Astronomical Society of India, 35, 87

\bibitem[{{Ranade} {et~al.}(2004){Ranade}, {Gupta}, {Ashok}, \&
  {Singh}}]{ranade04}
{Ranade}, A., {Gupta}, R., {Ashok}, N.~M., \& {Singh}, H.~P. 2004, Bulletin of
  the Astronomical Society of India, 32, 311

\bibitem[{{Rieke} \& {Lebofsky}(1985)}]{rieke85}
{Rieke}, G.~H. \& {Lebofsky}, M.~J. 1985, \apj, 288, 618

\bibitem[{{Rieke} {et~al.}(1989){Rieke}, {Rieke}, \& {Paul}}]{rieke89}
{Rieke}, G.~H., {Rieke}, M.~J., \& {Paul}, A.~E. 1989, \apj, 336, 752

\bibitem[{{Salpeter}(1955)}]{salpeter55}
{Salpeter}, E.~E. 1955, \apj, 121, 161

\bibitem[{{Skrutskie} {et~al.}(2006){Skrutskie}, {Cutri}, {Stiening},
  {Weinberg}, {Schneider}, {Carpenter}, {Beichman}, {Capps}, {Chester},
  {Elias}, {Huchra}, {Liebert}, {Lonsdale}, {Monet}, {Price}, {Seitzer},
  {Jarrett}, {Kirkpatrick}, {Gizis}, {Howard}, {Evans}, {Fowler}, {Fullmer},
  {Hurt}, {Light}, {Kopan}, {Marsh}, {McCallon}, {Tam}, {Van Dyk}, \&
  {Wheelock}}]{skrutskie06}
{Skrutskie}, M.~F., {Cutri}, R.~M., {Stiening}, R., {et~al.} 2006, \aj, 131,
  1163

\bibitem[{{Stetson}(1994)}]{stetson94}
{Stetson}, P.~B. 1994, \pasp, 106, 250

\bibitem[{{Tokunaga}(2000)}]{tokunaga00}
{Tokunaga}, A.~T. 2000, {Infrared Astronomy} ({Springer}), 143

\bibitem[{Vacca {et~al.}(2003)Vacca, Cushing, \& Rayner}]{vacca03}
Vacca, W.~D., Cushing, M.~C., \& Rayner, J.~T. 2003, \pasp, 115, 389

\bibitem[{{Wallace} \& {Hinkle}(1997)}]{wallacehinkle97}
{Wallace}, L. \& {Hinkle}, K. 1997, \apjs, 111, 445

\end{thebibliography}

\listofobjects

\end{document}